\documentclass[aps,floatfix,prd,twocolumn,preprintnumbers,
                superscriptaddress,groupedaddress]{revtex4}  

\usepackage{lineno}
\modulolinenumbers[5]
\usepackage{graphicx}  
\graphicspath{{./figures/}}
\usepackage{amsmath}   
\usepackage{amssymb}   
\usepackage{array}
\usepackage{slashed}
\usepackage[caption=false]{subfig}
\usepackage[colorlinks=true,
            urlcolor=blue,
            anchorcolor=blue,
            citecolor=blue,
            filecolor=blue,
            linkcolor=blue,
            menucolor=blue,
            linktocpage=true,
            pdfproducer=medialab,
            pdfa=true,
            bookmarks=false]{hyperref}
\usepackage{paralist}
\usepackage{multirow}
\usepackage{etoolbox}
\apptocmd{\thebibliography}{\raggedright}{}{}
\usepackage[capitalize]{cleveref}

\def\c{{\rm c}}

\def\s{{\rm s}}

\def\nl{\nonumber \\}

\makeatletter
 \def\@textbottom{\vskip \z@ \@plus 4pt}
 \let\@texttop\relax
\makeatother

\usepackage[normalem]{ulem}

\begin{document}
\preprint{ICPP-009}
\title{ Fingerprinting the Top quark FCNC via anomalous $Ztq$ couplings at the LHeC}
\author{Subhasish Behera}
\email{subhasish@iitg.ac.in}
\affiliation{Department of Physics,
	Indian Institute of Technology Guwahati, Assam 781039, India}
\author{Rashidul Islam}
\email{rislam@iitg.ac.in}
\affiliation{Department of Physics,
	Indian Institute of Technology Guwahati, Assam 781039, India}
\author{Mukesh Kumar}
\email{mukesh.kumar@cern.ch}
\affiliation{School of Physics and Institute for Collider Particle Physics,
	University of the Witwatersrand, Johannesburg, Wits 2050, South Africa}
\author{Poulose Poulose}
\email{poulose@iitg.ac.in}
\affiliation{Department of Physics,
	Indian Institute of Technology Guwahati, Assam 781039, India}
\author{Rafiqul Rahaman}
\email{rr13rs033@iiserkol.ac.in}
\affiliation{Department of Physical Sciences,\\
	Indian Institute of Science Education and Research Kolkata, Mohanpur 741246, India}


\begin{abstract}
 A study of the top quark \emph{Flavour Changing Neutral Current} (FCNC) through
 $Z$-boson has been performed in the proposed future $e^-p$ collider for the energy,
 $E_{e(p)} = 60~(7000)$~GeV. We considered an effective theory where the anomalous
 FCNC couplings are of vector and tensor nature.  The effect of these couplings is
 probed in the single top production along with the scattered electron. The polar
 angle $\theta$ of the electrons coming out of the primary vertex in association
 with the top quark polarization asymmetries constructed from the angular distribution
 of the secondary lepton arising from the top decay, allow to distinguish the Lorentz
 structure of the coupling. From a multi-parameter analysis, we obtain a reach of
 ${\cal O} (10^{-2})$ in the case of $Ztu$ and $Ztc$ couplings at an integrated luminosity
 of 2~ab$^{-1}$ at 95\% C.L.
\end{abstract}


\maketitle

%




\section{Introduction}
\label{sec:intro}
In the aftermath of the top quark discovery at the Tevatron, its properties like spin, charge, couplings with the other Standard Model (SM) particles etc. conform the SM values. Further, the Large Hadron Collider (LHC) measured its values very precisely~\cite{Group:2010ab,ATLAS:2014wva}. All these measurements over the years have established the top quark as the most interesting particle of the SM. Particularly, its mass around 173.1~GeV makes the top quark the heaviest among the SM particles and as a result allows it to decay much before the hadronization sets in. This behaviour single it out from other known quarks and gives us a probe of new physics~\cite{Dutta:2012ai, Boos:2015bta, Kroninger:2015oma,Coleppa:2017rgb}.

In the SM, the neutral current couples with the quarks as
\begin{gather}
 {\cal L}_{\rm NC}
 =
 - \frac{g}{2\c_W} \bar q \gamma^\mu (V - A\gamma_5) q Z_\mu ,
 \label{Lag:NC_SM}
\end{gather}
where $V = t_{3L} - 2Q\s^2_W$ and $A = t_{3L}$. Note that in the above Lagrangian the quarks are of the same flavour. The flavour changing neutral current (FCNC) is completely absent at the tree level. Not just that, even at the one loop level they are highly suppressed because of the GIM (Glashow-Iliopoulos-Maiani) mechanism~\cite{Glashow:1970gm}. For instance, the SM predictions for the branching fractions of FCNC processes like $t\to Z u(c)$ and $t \to \gamma u(c)$ are of the order of $10^{-17}(10^{-14})$ and $10^{-16}(10^{-14})$, respectively~\cite{Agashe:2013hma}. However, in beyond the SM (BSM) scenarios such suppression due to GIM mechanism can be relaxed, and one-loop diagrams mediated by new bosons may also contribute, yielding effective couplings of the orders of magnitude larger than those of the SM. We can express such an effective Lagrangian up to an energy scale $\Lambda$ as
\begin{gather}
 {\cal L}_{\rm eff}
 =
 {\cal L}^{(4)} + \frac{1}{\Lambda} {\cal L}^{(5)} + \frac{1}{\Lambda^2} {\cal L}^{(6)} + \cdots,
\end{gather}
where ${\cal L}^{(n)}$ consists of operators of dimension $n$ made of the SM fields obeying $SU(2)_L \otimes U(1)_Y$ gauge invariance. We can neglect the gauge invariant dimension 5 operator, ${\cal L}^{(5)}$ (responsible for Majorana masses of neutrinos), which has no relevance in the quark sector. However, ${\cal L}^{(4)}$ and ${\cal L}^{(6)}$ can contribute to the flavour changing interactions. ${\cal L}^{(4)}$ will consist of a vector current as shown in the \cref{Lag:NC_SM} albeit with dissimilar quark flavours. Similarly a tensorial flavour changing quark current will contribute to the ${\cal L}^{(6)}$.

Experiments performed earlier at the Tevatron and now at the LHC have failed to give us any interesting observation of FCNC. The bounds on such couplings from those experiments are very strong.

Here, we intend to study the possible BSM signature in the FCNC of the top quark sector in the proposed powerful high energy $e^-p$ collider, the {\em Large Hadron Electron Collider} (LHeC). With a choice of electron energy of $E_e = 60$~GeV, along with an available energy of LHC proton of $E_p = 7$~TeV, would provide a center of mass energy of $\sqrt{s} \approx 1.3$~TeV at the LHeC. Its design is such that the $e^-p$ and $pp$ colliders will operate simultaneously. Thus it would provide a cost effective alternative to all the future proposed colliders. Furthermore, The LHeC would gain advantage over the LHC or the Future Circular Collider for proton-proton (FCC-pp)~\cite{Kumar:2015kca} as
\begin{inparaenum}[(1)]
 \item initial states are asymmetric and hence backward and forward scattering can be disentangled,
 \item it provides a clean environment with suppressed backgrounds from strong interaction processes and free from issues like pile-ups, multiple interactions etc.,
 \item such machines are known for high precision measurements of the dynamical properties of the proton allowing simultaneous test of EW and QCD effects.
\end{inparaenum}
A detailed report on the physics and detector design concepts can be found in the Ref.~\cite{AbelleiraFernandez:2012cc}.

This article is arranged as follows. In \cref{sec:formalism}, we describe in detail the formalism used in this study. \cref{sec:fcnc} describes the FCNC effect in the top quark sector through the Effective Field Theory (EFT) approach and its experimental status. In \cref{sec:spin_matrix}, we detail the mechanism to construct asymmetries specific to top quark, whereas \cref{sec:el_assy} describes angular asymmetry of the primary electron. \cref{sec:analysis} gives the thorough analysis of the FCNC couplings from various aspects. \cref{sec:kinobs} gives the cut-based analysis and various distributions. \cref{sec:multi} gives the bounds arrived at form the multi-parameter analysis and likelihood analysis. Finally, we draw our inferences in \cref{sec:conc}.

\section{Formalism}
\label{sec:formalism}


\subsection{The Process}
\label{sec:fcnc}
The most general effective Lagrangian describing interactions of the top quark with light quarks $q=u,c$ and $Z$ boson allowing FCNC processes can be given by~\cite{AguilarSaavedra:2008zc},
\begin{multline}
 {\cal L}_{Ztq}
 =
 - \frac{g}{2c_W}\bar{q}\gamma^{\mu}(X_{qt}^LP_L + X_{qt}^RP_R)t~Z_{\mu}
 \\
 - \left[ \frac{g}{2c_W}\bar{q}\frac{i \sigma^{\mu \nu}(p_t-p_q)_{\nu}}{\Lambda}
    (\kappa_{qt}^LP_L + \kappa_{qt}^RP_R)t~Z_{\mu} + {\rm h.c} \right],
 \label{eq:Lag}
\end{multline}
where $(p_t - p_q)$ is the momentum transfer between the quarks in the process, and $\Lambda$ is the cut-off scale, which we set as the top quark mass ($\Lambda = m_t$). The vector couplings are denoted by $X_{qt}^{L,R}$ and the tensor couplings by $\kappa_{qt}^{L,R}$. The choice of scale $\Lambda$ at $m_t$ is motivated from the minimum energy required to produce at least one {\em on-shell} top quark.  As we can see the vector couplings are independent of $\Lambda$, whereas its effect on the tensor couplings can be derived easily by using the substitution $\kappa_{qt}^{L,R} \to \kappa_{qt}^{L,R}\,m_t/\Lambda$.  Coming to the present constraints on the above couplings, the CMS collaboration of the LHC has performed a search for single top quark production with $Z$-boson events with 5~fb$^{-1}$ data at $\sqrt{s}=7$~TeV~\cite{CMS:2013nea}. Subsequently, from the non-observance of FCNC they put the following bounds: $\sqrt{2} \kappa^L_{ut}/\Lambda < 0.45$~TeV$^{-1}$ corresponding to ${\rm BR}(t\to Zu) \leq 0.51\%$, and $\sqrt{2} \kappa^L_{ct}/\Lambda < 2.27$~TeV$^{-1}$ corresponding to ${\rm BR}(t\to Zc) \leq 11.40\%$. A similar search for FCNC in top quark decay $t\to Zq$ has been performed by the CMS corresponding to a luminosity of 19.7~fb$^{-1}$ at $\sqrt{s}=$8~TeV from the decay chain $t\bar{t}\to Zq + Wb$, where both vector boson decay leptonically, producing a final state with three leptons (electrons or muons)~\cite{Chatrchyan:2013nwa}, excluding ${\rm BR}(t \to Zq)>0.05\%$ at the 95\% confidence level. The latest ATLAS search at $\sqrt{s}=13$~TeV~\cite{Aaboud:2018nyl} with a luminosity of 36.1~fb$^{-1}$ sets ${\rm BR}(t\to Zu(c))<1.7(2.4)\times 10^{-4}$. The event considered for investigation was $p p \to t\bar{t},(t\to Zq(u,c),\bar{t}\to W^+b)$ with both $Z,W$-bosons decay leptonically.  Projected reach of these BR's at the high luminosity LHC with 3 ab$^{-1}$ luminosity (HL-LHC) are  $2.5-5.5 \times 10^{-5}$ \cite{ATL-PHYS-PUB-2016-019}. In future high energy $e^+e^-$ collider the $Ztq$ effective couplings can be excluded up to ${\cal O}(10^{-5})$ for 300~fb$^{-1}$~\cite{Khanpour:2014xla}. Recently the authors of Ref.~\cite{Cakir:2018ruj} performed a similar study for $e^-p$ scenario.

\begin{figure}[!ht]
  \centering
  \includegraphics[trim=0 0 0 0,clip,width=0.7\linewidth]{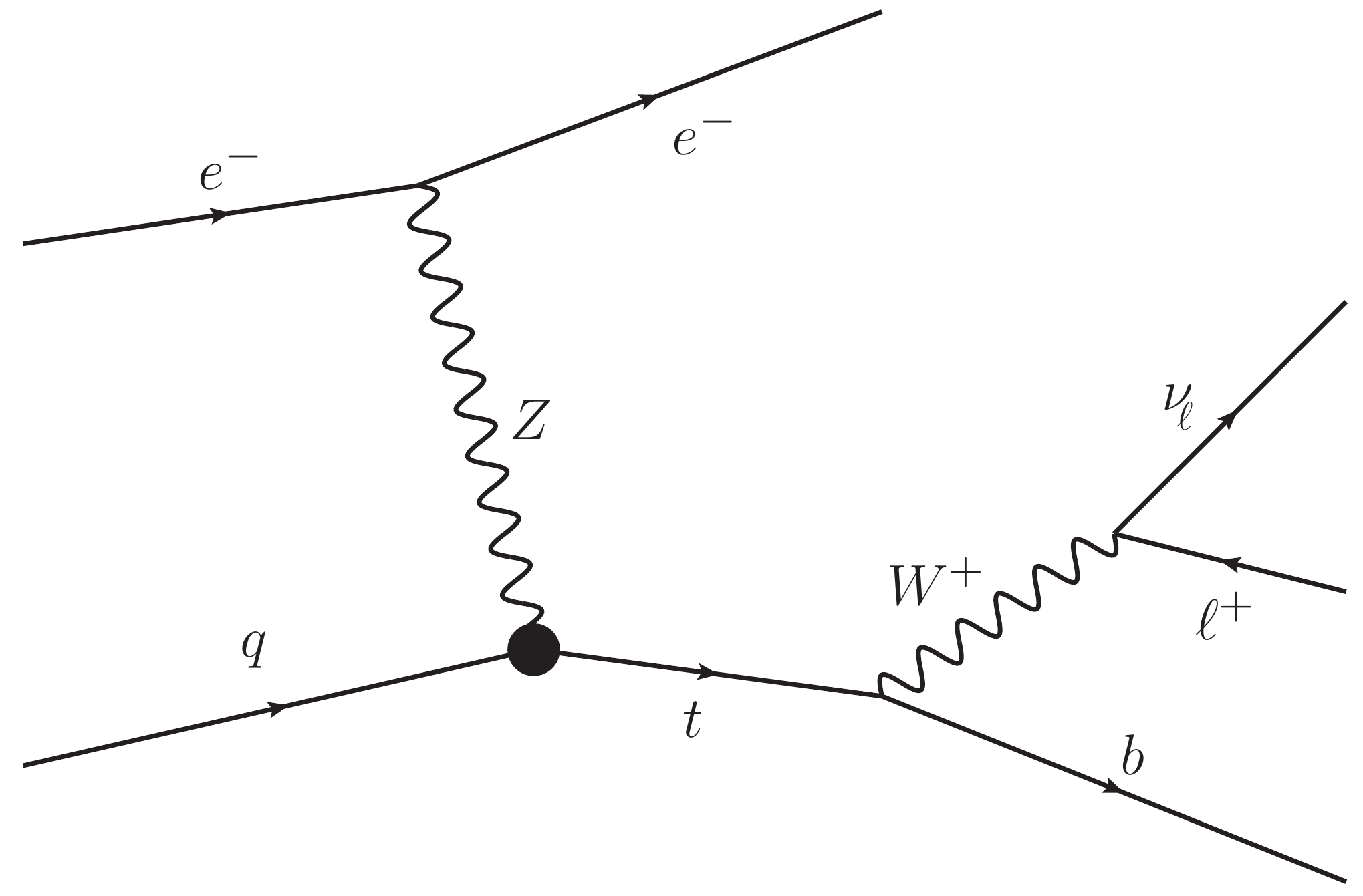}
  \caption{Signal processes: The $Ztq$ anomalous vertex at production channel of top
           and decay of top via SM coupling vertex only. The final state charge
           lepton is $\ell^+$ in our study to make proper distinction between
           charged lepton ($e^-$) coming out of the $e^-Ze^-$-primary vertex and
           decay of the top it self.}
  \label{fig:FCNC_signal}
\end{figure}

The single top-quark production process at $e^-p$ collider (a detailed study through charged-current top-quark production in this environment is performed in Refs.~\cite{Dutta:2013mva, Kumar:2015jna}) enabled by these interactions is a $t$-channel exchange of $Z$ boson coupling the quarks with the leptons, $e^-p \to e^- t, (t \to W^+ b, W^+ \to \ell^+ \nu_\ell)$, where $\ell=e,~\mu$, the Feynman diagram of which is shown in \cref{fig:FCNC_signal}. We consider the leptonic decay of the top quark keeping in mind the spin-correlation study and the top polarization asymmetries that might be useful in the investigation of the $Ztq$ anomalous couplings.
Note that, there is no SM analogue of this process, thus we expect the backgrounds can be reduced without difficulty.

A few comments are in order before we get into the details of the analyses. 
\begin{enumerate}
 \item At the first look, the scattered electron (as opposed to the electron or muon produced in the $W$ decay) is a spectator, not connected with the top quark production vertex. However, notice that the $Z$ boson coupling to the top quark is sensitive to its state of polarization. This in turn can reflect in the polarisation state, and consequently the angular distribution of the scattered electron. We shall exploit this situation in our analyses, and construct observables based on the kinematic distribution of the scattered electron.
 \item The second observation is that the top quark polarisation is directly affected by the nature of the coupling. That is, whether it is a vector coupling or a tensor coupling, and whether it couples to the left-handed or the right-handed quarks. It is well known that the spin information of the top quark will be carried forward to the decay products, and will be reflected in the angular and energy distribution of the secondary leptons. We shall make use of this fact in constructing multiple observables, a combination of which could discriminate the type of $Ztq$ couplings.
\end{enumerate}
In the following section we shall elaborate on the top quark spin analysis and various asymmetries making use of this information, which would be employed in the study.

\subsection{Polarization of the top quark}
\label{sec:spin_matrix}
In this section we discuss the formalism that could be employed to extract the polarization information of the top quark through suitably constructed observables. For details of the formalism one may consult Ref.~\cite{Godbole:2006tq, Boudjema:2009fz,Rindani:2011pk}. As explained in the previous section, the motivation for the spin analysis of top quark comes from the fact that the angular distributions of top quark decay products give access to the Lorentz structure of the production vertex through the information of top quark polarisation.

In the Narrow Width Approximation (NWA), the invariant amplitude square of the full process ($eq\rightarrow et \rightarrow eb\ell\nu$) can be written as a product of the production and decay density matrices in the helicity basis of the top quark as
 \begin{equation}\label{eq:3}
 {|{\cal M}|^2}
 =
 \frac{\pi\delta(p_t^2 -m_t^2)}{\Gamma_t m_t}
  \sum_{\lambda,\lambda'}\rho(\lambda,\lambda')\Gamma(\lambda,\lambda')
\end{equation}
where $p_t$ is the momentum and $\Gamma_t$ is the total width of the top quark, with the summation considered over the helicity indices of the top quark. The production and decay density matrices are given in terms of the corresponding amplitudes as $\rho(\lambda,\lambda') = {\cal M}_P(\lambda)\,{\cal M}_P^*(\lambda')$ and $\Gamma(\lambda,\lambda') = {\cal M}_\Gamma(\lambda)\,{\cal M}_\Gamma^*(\lambda')$, respectively. The top quark on-shell condition in the NWA allows one to define the normalised production density matrix of the top quark as
\begin{align}
 \sigma(\lambda,\lambda')
 =
 \frac{1}{\sigma_{\rm prod}}\int \rho(\lambda,\lambda') d\Omega_t,
\end{align}
where $d\Omega_t$ is the differential solid angle of top quark produced (for details, please refer to~\cite{Boudjema:2009fz}) and $\sigma_{\rm prod}$ is the total production cross section. 
%
For convenience, we define polarisation vector ${\bf P}=(P_x,P_y,P_z)$ so that
%
%
\begin{align}
\begin{aligned}
\sigma(+,+)&=\frac{1}{2}(1+P_z), \\
\sigma(-,-)&=\frac{1}{2}(1-P_z), \\
\sigma(+,-)&=\frac{1}{2}(P_x+iP_y) \\
\sigma(-,+)&=\frac{1}{2}(P_x-iP_y).
\end{aligned}
\end{align}
The normalized decay density matrix elements for the process $t\to W^+ b\to b \ell^+ \nu_\ell$ may be written in terms of the polar ($\theta_\ell$) and azimuthal ($\phi_\ell$) angles of the secondary lepton in the top rest frame as~\cite{Boudjema:2009fz},
%
%
\begin{align}
\begin{aligned}
\Gamma(+,+)&=\frac{1}{2} (1+ \cos \theta_\ell) , \\ 
\Gamma(-,-)&=\frac{1}{2} (1- \cos \theta_\ell) , \\
\Gamma(+,-)&=\frac{1}{2} \sin \theta_\ell e^{i\phi_\ell}, \\ 
\Gamma(-,+)&=\frac{1}{2} \sin \theta_\ell e^{-i\phi_\ell}.
\end{aligned}
\label{decay_matrix}
\end{align}
Here the polar angle is measured with respect to the top quark boost direction, and the top production plane is taken as the $x$-$z$ plane. These choices of reference do not cost us generality of the analysis as shown in Ref.~\cite{Godbole:2006tq}. The differential cross section for the complete process in terms of the top quark polarisation vector and the polar and azimuthal angle of the secondary lepton in the rest frame of the top quark, can now be written as

\begin{align}
  \frac{1}{\sigma_{\rm tot}}\frac{d\sigma}{d\Omega_\ell}
  =
  \frac{1}{4\pi} \Big(1+&\,P_z \cos\theta_\ell + P_x \sin\theta_\ell \cos\phi_\ell
  \nonumber\\
  +&\, P_y \sin\theta_\ell \sin\phi_\ell \Big),
\end{align}
where $\sigma_{\rm tot}=\sigma_{\rm prod}\times {\rm BR}(t\rightarrow b\ell \nu)$.
This enables one to define angular asymmetries of the secondary leptons, and connect those directly to the top quark polarisation. The following three asymmetries of this kind~\cite{Godbole:2006tq}, two defined in terms of the azimuthal angle, and one in terms of the polar angle of the decay lepton, are used in the subsequent study.
\begin{align}
\begin{aligned}
 A_x \equiv\,&
 \frac{1}{\sigma_{\rm tot}}\bigg[
  \int_{-\frac{\pi}{2}}^{\frac{\pi}{2}} \!\! d\phi_\ell \frac{d\sigma}{d\phi_\ell}
 - \int_{\frac{\pi}{2}}^{\frac{3\pi}{2}} \!\! d\phi_\ell \frac{d\sigma}{d\phi_\ell}
 \bigg] =\frac{1}{2} P_x , \\
 A_y \equiv\,&
 \frac{1}{\sigma_{\rm tot}}\bigg[
  \int_{0}^{\pi} \!\! d\phi_\ell \frac{d\sigma}{d\phi_\ell}
 - \int_{\pi}^{2\pi} \!\! d\phi_\ell \frac{d\sigma}{d\phi_\ell}
 \bigg] =\frac{1}{2} P_y , \\
 A_z \equiv\,&
 \frac{1}{\sigma_{\rm tot}}\bigg[
  \int_{0}^{1} \!\! d\c_{\theta_\ell} \frac{d\sigma}{d\c_{\theta_\ell}}
 - \int_{-1}^{0} \!\! d\c_{\theta_\ell} \frac{d\sigma}{d\c_{\theta_\ell}}
 \bigg]=\frac{1}{2} P_z.
\end{aligned}
 \label{eq_asym}
\end{align}
Note that the angles in the above asymmetries are defined in the rest frame of the top quark, and thus require full reconstruction of the top quark momentum. In the present case this leads to the following relations between the components of the missing momentum (neutrino in this case) denoted by $p_{x\nu},~p_{y\nu},~p_{z\nu}$, and those of the visible final particles.
\begin{eqnarray}
p_{x\nu} &=& - \sum_{k=e,\ell,b} p_{x k},
 \qquad p_{y\nu} = - \sum_{k=e,\ell,b} p_{y k}, \nonumber\\
(p_{z\nu})_\pm
&=&
\frac{1}{p_{T\ell}^2}
 \Big[ \beta p_{z\ell} \mp E_\ell \sqrt{\beta^2 - p_{T\nu}^2 p_{T\ell}^2} \Big],
\end{eqnarray}
where $\beta = \frac{m_W^2}{2} + p_{x\ell} p_{x\nu} + p_{y\ell} p_{y\nu}$ and $p_{Ti}^2=p_{xi}^2+p_{yi}^2$. Out of the above two solutions for $p_{z\nu}$, the one for which $|\sum_j p_j^2 - m_t^2|$ is minimum, where $p_j$ is the four momentum of the corresponding particle, with $j=\ell,b,\nu$, will be considered as the correct choice for the $z$-component. The missing momentum thus obtained is used to reconstruct the top quark momentum. The reader may note that the accuracy of this reconstruction of the top momentum depends on the precise measurements of the lepton and jet momenta and energy. In our numerical analysis we take into account of all these effects through an assumed systematic uncertainty.

\subsection{Angular asymmetry of the recoiled electron}
\label{sec:el_assy}
As mentioned in the introduction, the angular distribution of the scattered electron is indirectly sensitive to the Lorentz structure of the $Ztq$ interaction. Exploiting this unique feature of LHeC, we define forward-backward asymmetry (in the lab frame) of the $e^-$ coming out of the primary vertex
\begin{align}
 A_{e}^{FB}
 =
 \frac{\sigma(\cos \theta_e > 0)-\sigma(\cos \theta_e < 0)}
      {\sigma(\cos \theta_e > 0)+\sigma(\cos \theta_e < 0)}.
 \label{a_e}
\end{align}
Notice that the other lepton coming from the decay of the top quark is positively charged, and we assume identifying the charge of the leptons. 

In the rest of the article we shall demonstrate that these asymmetries along with the cross section itself could be effectively employed to identify and distinguish the $Ztq$ couplings.

\section{Simulation and Analysis}
\label{sec:analysis}
We perform the analyses with events generated using Monte Carlo event generator {\sc MadGraph5}~\cite{Alwall:2014hca}, using the model for signal events implemented using {\sc FeynRules}~\citep{Alloul:2013bka} package. Showering, fragmentation and hadronization are performed with customized {\sc Pythia-PGS}~\cite{Sjostrand:2006za}. The events thus generated are passed through {\sc FastJet}\cite{Cacciari:2011ma} for jet formation within $\Delta R = 0.4$, and {\sc Delphes}~\cite{deFavereau:2013fsa} to emulate the detector effects where an appropriately customised detector card being used. To generate the signal events CTEQ6L1 PDF set is used fixing the factorization and renormalization scale to be the threshold value of the top quark mass, $\mu = \mu_F = \mu_R = m_t$. However, for all background events these scales are set to be of a dynamical scale based on events. As preliminary selection criteria at the event generation level we considered $p_T > 10$~GeV and $|\eta| < 5$ for all light jets, $b$-jets and leptons, $\slashed E_T > 10$~GeV and the separation of $\Delta R_{ij} > 0.4$ between all possible jets and leptons or photons. The cross section of signal events for different FCNC couplings taken one at a time are given in Table~\ref{tab:cutflow_sig} for initial electron beam polarization of $-80\%$. The corresponding cross section for the other values of $e^-$ beam polarization, $P_e$ can be obtained using the formula $\sigma_{\rm pol} = \sigma_{\rm unpol} \times (1 - P_e)$. The main background processes and the corresponding cross sections are given in Table~\ref{tab:cutflow_bkg}. Notice that there are no background mimicking the same final state at the parton level. However, we considered all probable cases that could arise due to misidentification of particles leading to background emerging at the detector level. These include $(i)$ charged current processes like $ep\rightarrow e jW\rightarrow ejjj,~ep\rightarrow eWj\rightarrow ej\ell\nu,~ep\rightarrow \nu_eWb\rightarrow \nu_e\ell\nu b,~$, $(ii)$ neutral current processes like $ep\rightarrow eZb \rightarrow ebbb,~ep\rightarrow eZb\rightarrow eb\ell \ell,~ ep\rightarrow eZb\rightarrow ebjj$ and $(iii)$ photo-processes like $p\gamma \rightarrow \ell\ell b,~p\gamma\rightarrow \nu\nu \ell\ell b$.
\begin{table}[!ht]
\centering
\begin{tabular}{c|r|r|r|r} \hline
 \multirow{2}*{Coupling}
                 &\multicolumn{4}{c}{Cross section $\sigma$ in fb for $P_e = - 80\%$}
                                                                                   \\ \cline{2-5}
 $g_{Ztq}$                & Basic Cuts & $N_{e^-}=1$ & $N_{e^-,b}=1$ & $N_{e^-,b,\ell^+}=1$ \\ \hline\hline
 $X_{ut}^L$      & 1957.57    & 1763.82     & 799.65        & 745.57               \\ \hline
 $X_{ut}^R$      & 1642.47    & 1485.97     & 706.09        & 629.54               \\ \hline
 $\kappa_{ut}^L$ &  706.77    &  636.65     & 304.56        & 279.13               \\ \hline
 $\kappa_{ut}^R$ & 1038.68    &  933.47     & 474.90        & 427.77               \\ \hline
 $X_{ct}^L$      &  136.76    &  122.54     &  66.90        &  62.84               \\ \hline
 $X_{ct}^R$      &  103.82    &   93.05     &  51.26        &  47.65               \\ \hline
 $\kappa_{ct}^L$ &   26.37    &   23.65     &  12.96        &  12.09               \\ \hline
 $\kappa_{ct}^R$ &   60.00    &   53.33     &  29.70        &  27.45               \\ \hline
\end{tabular}
\caption{The signal cross sections for different anomalous $Ztq, (q=u,c)$ couplings, $g_{Ztq}$,
         at beam energies, $E_{e^-} = 60$ GeV and $E_p = 7$ TeV for polarized
         electron beam of $-80\%$. The cross section can be obtained from the above table
         as $\sigma = |g_{Ztq}|^2 [ \sigma (pe^- \to e^- t) \times {\rm BR}(t \to \ell^+
         + b\text{-tagged jet} + \slashed E_T) ]$. In the case of tensor couplings, the
         scale $\Lambda = m_t$.}
\label{tab:cutflow_sig} 
\end{table}
For further selection of events, in addition to the basic cuts (BC) employed at the generation level preliminary selection, we demand that the event contain exactly one $e^-$, one $b$-jet and one $\ell^+$. The signal and background cross sections, for $-80$\% polarization of the initial $e^-$ beam, after this selection is presented in \cref{tab:cutflow_sig} and \cref{tab:cutflow_bkg}, respectively. A $p_T$ dependent $b$-tagging efficiency of about 70\% is considered as expected. Most of the backgrounds are eliminated at this stage, with the remaining background cross section totalling to about 2.3 fb. We consider this remaining background throughout our analysis.

\begin{table}[!ht]
 \centering
 \resizebox{\linewidth}{!}{
 \begin{tabular}{c|r|r|r|c} \hline
 \multirow{2}*{Bkg processes}
                 &\multicolumn{4}{c}{Cross section $\sigma$ in fb for $P_e = - 80\%$}
                                                                                   \\ \cline{2-5}
                 & Basic Cuts & $N_{e^-}=1$ & $N_{e^-,b}=1$ & $N_{e^-,b,\ell^+}=1$ \\ \hline\hline 
   Charged Current int.  &67441.71&477.50&154.26&0.63 \\ \hline
   Neutral Current int.    &339289.52&293361.77&8110.97&0.00 \\ \hline
   Photo Production int. &29091.31&1339.20&435.45&1.64 \\ \hline
 \end{tabular}
 }
 \caption{The SM background cross sections at beam energies, $E_{e^-} = 60$~GeV
		and $E_p = 7$~TeV for electron polarization $-80\%$.}
 \label{tab:cutflow_bkg}
\end{table}

\subsection{Asymmetries}
\label{sec:kinobs}
\begin{figure}[!ht]
 \centering
 \includegraphics[trim=0 0 0 0,clip,width=\linewidth]{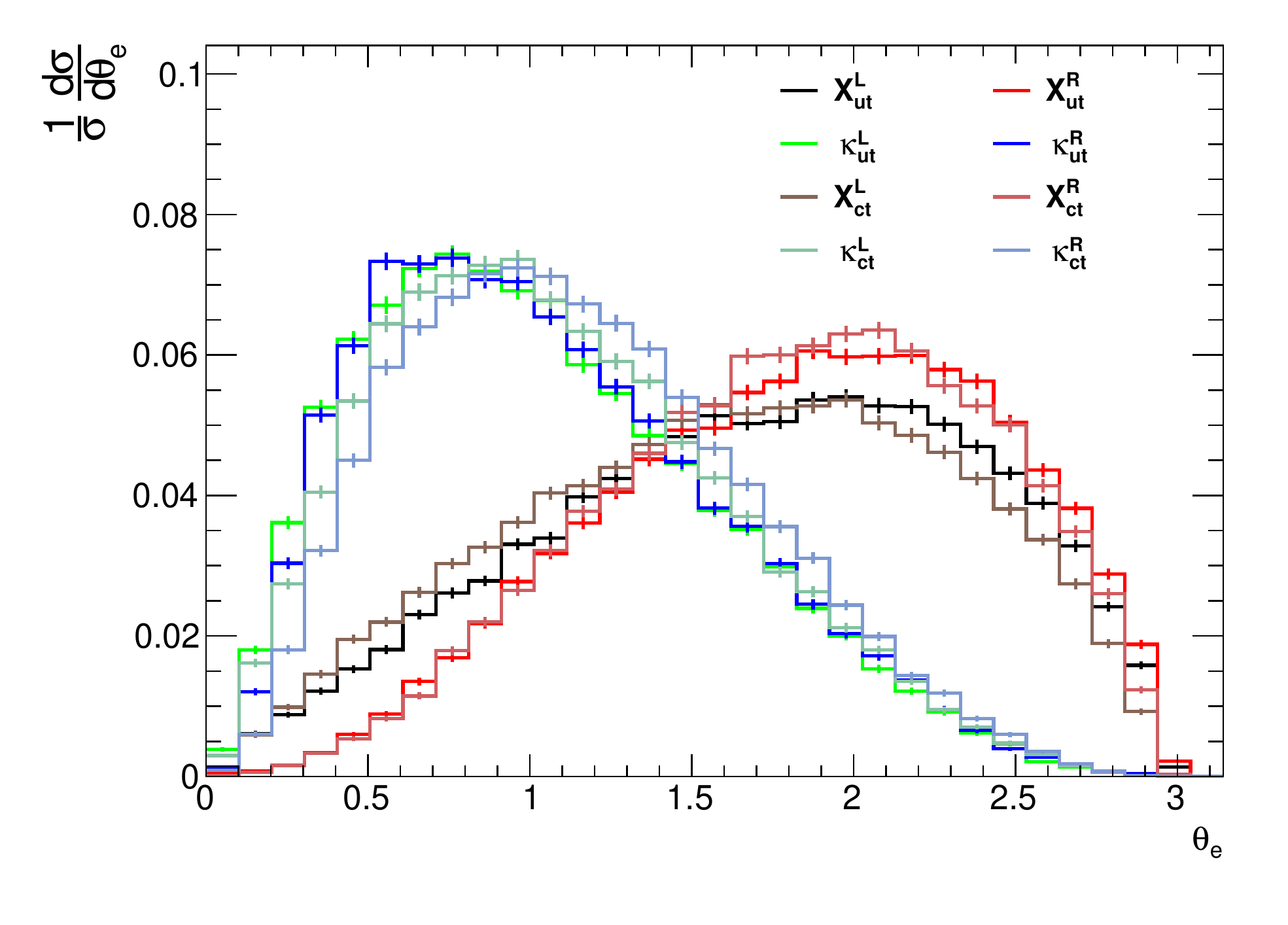}
 \caption{Polar angle distributions of the scattered $e^-$ in
          the final state for $-80\%$ polarization of electron beam. Note that $X$
          and $\kappa$ denote the vector and tensor couplings respectively.}
 \label{fig:ele_theta_dist}
\end{figure}
In this section we study the asymmetries as defined in \cref{eq_asym,a_e}. To study these asymmetries we plot the angular distribution of the scattered electron as shown in \cref{fig:ele_theta_dist}. Here, the forward direction is defined as the direction of the proton beam. Notice that all the vector couplings prefer electrons coming opposite to the proton direction (or along the incoming electron beam direction), which corresponds to smaller fraction of backward scattering. On the other hand, all the tensor couplings exhibit large backward scattering of the electron, indicating the requirement of larger momentum transfer. This is expected from the nature of the coupling, which is proportional to the momentum transfer. Moreover, the case of vector couplings allow slight discrimination between the left- and right-handed couplings. We may caution the reader that this distinguishability is limited by statistics, and perhaps not possible for $Ztc$ couplings even with very large luminosity. However, the possibility is quite realistic in the case of $Ztu$ couplings. This asymmetry, along with the top quark polarisation asymmetry are given in \cref{asym_discrim}, assuming that only one type of coupling is present (in each case). Note that the asymmetries (when only one type of couplings is present) are independent of the actual value of the coupling, as the dependence gets cancelled between its numerator and the denominator. On the other hand, if more than one type of couplings contribute, then this cancellation does not occur due to their interference, and the asymmetry depends on the actual value of the coupling. We shall discuss the multi-parameter case and the analyses in the next subsection. The correlated dependence of the $A_e^{FB}$ asymmetry on the left/right-handed couplings and the beam polarisation is quite clearly indicated as well in \cref{asym_discrim}. Larger asymmetries are present when the electron beam polarisation and the handedness of the couplings are opposite in the case of vector couplings. 

\begin{table}[!ht]
	\resizebox{\linewidth}{!}{
	\begin{tabular}{c|c|c|c} 
		\multicolumn{4}{c}{Left-polarised $e$-beam}\\ \hline
		$A_x$&$A_z$                & $A_{e}^{FB}$   &{\small Coupling} \\ \hline\hline
		  $-0.16$&$-0.43$&$-0.18$&$X^L$\\
		  $-0.17$&$-0.46$&$+0.63$&$\kappa^L$\\
		 $+0.07$&$+0.32$&$-0.33$&$X^R$\\
		 $+0.04$&$+0.37$&$+0.65$&$\kappa^R$\\\hline
	\end{tabular}
\hskip 5mm
	\begin{tabular}{c|c|c|c} 
		\multicolumn{4}{c}{Right-polarised $e$-beam}\\ \hline
		$A_x$&$A_z$                & $A_{e}^{FB}$   &{\small Coupling} \\ \hline\hline 
	    	$-0.06$&$-0.43$&  $ -0.34$&$X^L$\\
		    $-0.01$&$-0.46$&    $+0.64$&$\kappa^L$\\
		   $+0.16$&$+0.32$&  $ -0.17$&$X^R$\\
		   $+0.16$&$+0.37$&   $ +0.65$&$\kappa^R$\\ \hline
	\end{tabular}
}
	\centering
	\begin{tabular}{c|c|c|c} 
		\multicolumn{4}{c}{Unpolarised $e$-beam}\\ \hline
		$A_x$&$A_z$                & $A_{e}^{FB}$   &{\small Coupling} \\ \hline\hline 
		$-0.12$&$-0.43$&   $-0.24$&$X^L$\\
		$-0.12$&$-0.46$&    $+0.64$&$\kappa^L$\\
		$+0.11$&$+0.32$&  $ -0.26$&$X^R$\\
		$+0.08$&$+0.36$&    $+0.65$&$\kappa^R$\\ \hline
	\end{tabular}
	\caption{Asymmetries for one fixed value of coupling at a time. It shows the distinction
        among $X_{qt}^L$, $X_{qt}^R$, $\kappa_{qt}^L$ and $\kappa_{qt}^R$ by just looking at
        the sign of $A_z$ (Top quark rest frame observable) and $A_{e}^{FB}$ (Lab frame
		observable) as shown in \cref{eq_asym,a_e}.}
	\label{asym_discrim}
\end{table}

Among the top polarisation asymmetries, $A_y=P_y$ is identically zero owing to the $CP$ symmetry of the interactions considered. When only one type of coupling is present both $\sigma(+,+)$ and $\sigma(-,-)$ are expected to be the same leading to $A_z=\frac{P_z}{2}=\frac{1}{2}$. However, we notice a small deviation from the value of $\frac{1}{2}$ in the \cref{asym_discrim} which might be due to the effects like that of the detector simulation, particle identification efficiency, etc. As expected, the left-handed couplings and right-handed couplings give rise to negative and positive asymmetries, respectively, thus giving a handle to discriminate the type of couplings. With unpolarised electron beam, $A_x$ is close to 10\% for all cases of couplings, except for left-handed tensor couplings for which it is negligible. With the beam polarisation, this features the distinguishing ability, with the asymmetry vanishing for the opposite combination of polarisation. That is, $A_x$ is negligible for right-handed couplings, when the electron beam is left-polarised, and vice versa. Thus, a combination of the asymmetries measured with left-polarised, right-polarised and unpolarised electron beam provide clear indication of the type of the coupling present. 

\subsection{Multi-parameter Analysis}
\label{sec:multi}
Going beyond the single parameter case, we shall now consider simultaneous presence of more than one parameter and the reach on their values that may be obtained at an $e^-p$ collider through single top production being considered in this discussion. We shall restrict to the case when either of $u$ or $c$ quark is considered at a time. The cross section can be written as a second order polynomial in the relevant parameters, as follows 
\begin{eqnarray}
\sigma_{\rm tot} (fb) &= &745.57 ~{X_{ut}^L}^2 + 629.54 ~{X_{ut}^R}^2+ 279.13 ~{\kappa_{ut}^L}^2 \nl 
&&+ 427.77 ~ {\kappa_{ut}^R}^2 - 7.96 ~X_{ut}^L~\kappa_{ut}^R + 0.97 ~X_{ut}^R~\kappa_{ut}^L \nl
&&+ 62.84 ~{X_{ct}^L}^2 + 47.65 ~{X_{ct}^R}^2 + 12.09 ~{\kappa_{ct}^L}^2 \nl
 &&+ 27.45 ~{\kappa_{ct}^R}^2 - 0.91 ~X_{ct}^L~\kappa_{ct}^R - 2.48 ~X_{ct}^R~\kappa_{ct}^L \nl
 \label{eq_func_cs}
\end{eqnarray}
The normalised top-polarisation asymmetries may similarly be written as
\begin{equation}
 A_i = \frac{A_i^N}{\sigma_{\rm tot}}, \qquad i = x,z, e(FB),
\end{equation}
where $A_i^N\times{\cal L}$ , with ${\cal L}$ denoting the integrated luminosity, will give the asymmetric number of events. $A_i^N$ can also be expressed as a polynomial function of the coupling parameters as given below. The coefficients in this case are obtained by a numerical fit.
\begin{eqnarray}
 A_{x}^N &= &   -119.90~{X_{ut}^L}^2 
                    +    44.02~{X_{ut}^R}^2 
                     - 45.68~{\kappa_{ut}^L}^2  \nl 
  &&             + 16.86 ~ {\kappa_{ut}^R}^2 
                    +    3.89~X_{ut}^L~\kappa_{ut}^R 
                    - 3.20 ~X_{ut}^R~\kappa_{ut}^L \nl
  &&             - 13.85~{X_{ct}^L}^2  
                    + 1.08 ~{X_{ct}^R}^2 
                    - 2.45~{\kappa_{ct}^L}^2\nl
  &&             - 0.36~{\kappa_{ct}^R}^2 
                    + 1.61~X_{ct}^L~\kappa_{ct}^R 
                    -3.61 ~X_{ct}^R~\kappa_{ct}^L\nl
\end{eqnarray}
\begin{eqnarray}  
  A_{z}^N &= & 
        -320.27~{X_{ut}^L}^2 
  + 199.36~{X_{ut}^R}^2 
  - 125.51~{\kappa_{ut}^L}^2  \nl 
  && + 151.95~ {\kappa_{ut}^R}^2 
  +    5.36~X_{ut}^L~\kappa_{ut}^R 
  + 11.65~X_{ut}^R~\kappa_{ut}^L \nl
  &&             - 25.97~{X_{ct}^L}^2  
  + 14.28~{X_{ct}^R}^2 
 - 6.18~{\kappa_{ct}^L}^2\nl
  &&            + 7.76 ~{\kappa_{ct}^R}^2 
  + 3.57~X_{ct}^L~\kappa_{ct}^R 
  +      8.41~X_{ct}^R~\kappa_{ct}^L\nl
\end{eqnarray}
\begin{eqnarray}
 {A_{e}^{FB}}^N &=&
        -134.93~{X_{ut}^L}^2 
 - 206.87~{X_{ut}^R}^2 
+ 170.14~{\kappa_{ut}^L}^2  \nl 
 && + 269.34~ {\kappa_{ut}^R}^2 
 -2.85 ~X_{ut}^L~\kappa_{ut}^R 
 + 3.57 ~X_{ut}^R~\kappa_{ut}^L \nl
 &&             -    6.85 ~{X_{ct}^L}^2  
 -     15.81 ~{X_{ct}^R}^2 
 + 6.30~{\kappa_{ct}^L}^2\nl
 &&           + 16.00~{\kappa_{ct}^R}^2 
 + 2.81~X_{ct}^L~\kappa_{ct}^R 
 -2.37~X_{ct}^R~\kappa_{ct}^L\nl
 \label{eq_func_as}
\end{eqnarray}
We tried to include all the possible terms irrespective of their significance.  Terms with smaller coefficients are less significant. This  means that the quadratic couplings give the dominant contributions. Similarly, the tensor couplings are sub-leading compared to the vector couplings, when considered together. 
In the rest of this section, we shall make use of the above information on the cross section and the asymmetries to obtain the reach of the $e^-p$ collider in extracting the anomalous FCNC couplings. Apart from the single parameter analysis, where one assumes that only one of the couplings is present at a time, we shall also investigate the possibilities when more than one couplings present simultaneously. For single and two parameter cases we shall employ $\chi^2$ analysis, whereas considering simultaneous presence of all couplings, we shall perform a likelihood analysis to extract the information regarding reach of the collider at 2~ab$^{-1}$ luminosity.

\subsubsection{$\chi^2$ Analysis}
\begin{figure}[!ht]
	\centering
	\includegraphics[trim=0 0 0 0,clip,width=0.48\linewidth]{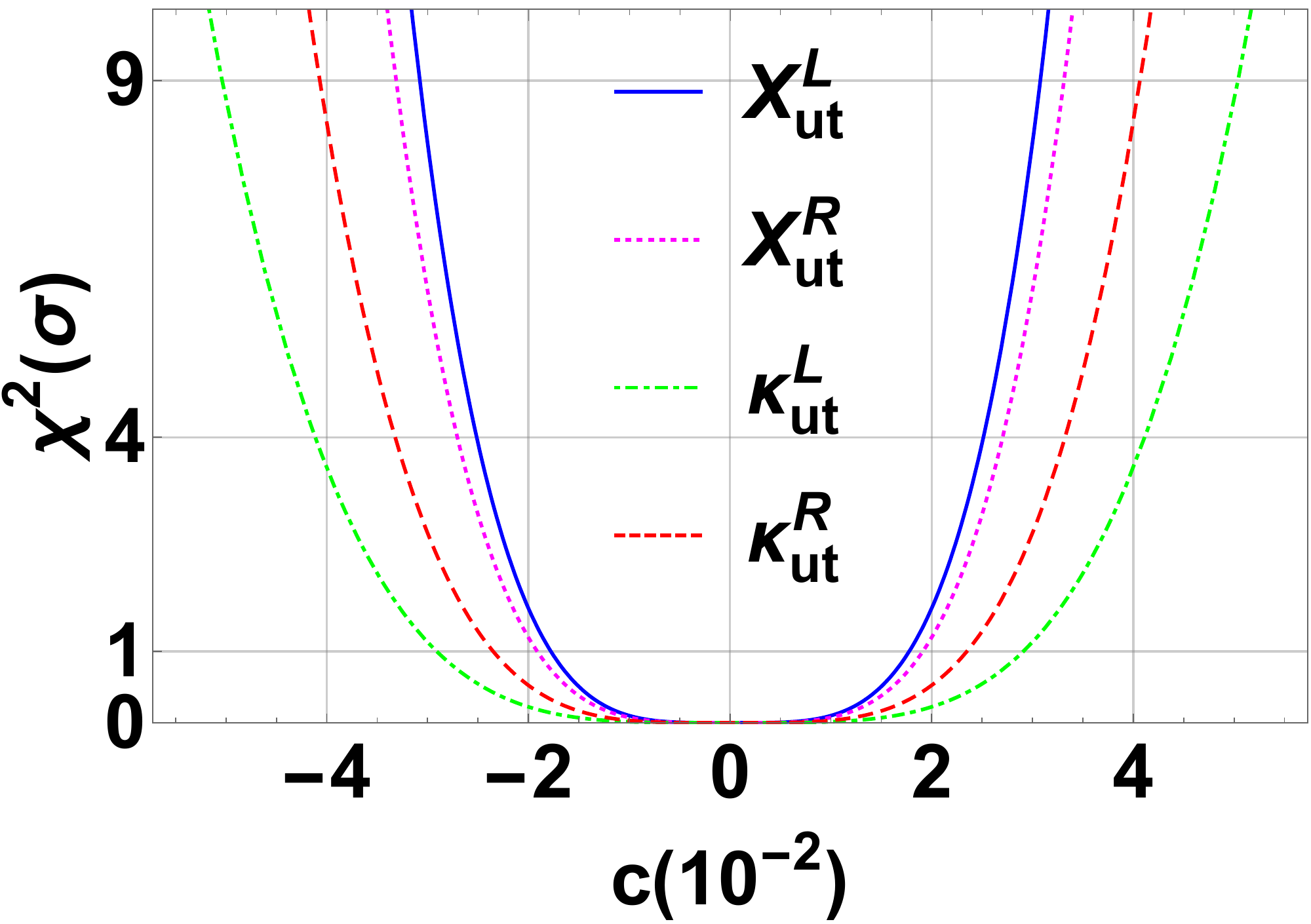} 
	\includegraphics[trim=0 0 0 0,clip,width=0.48\linewidth]{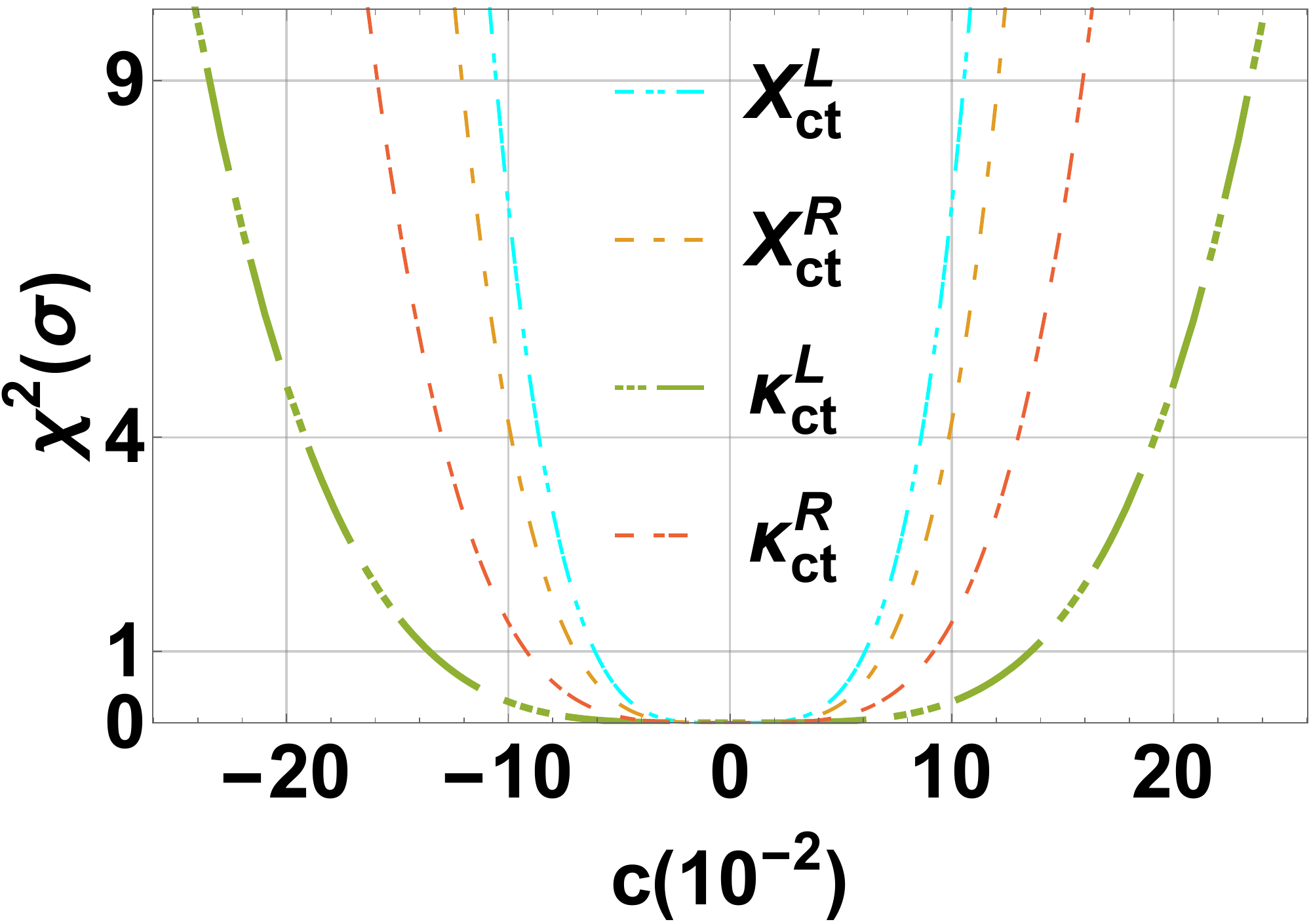}
	\caption{Single parameter reach with an integrated luminosity of 2~ab$^{-1}$.}
	\label{chi_one}
\end{figure}

\begin{figure}[!ht]
	\centering
	\includegraphics[trim=0 0 0 0,clip,width=0.48\linewidth]{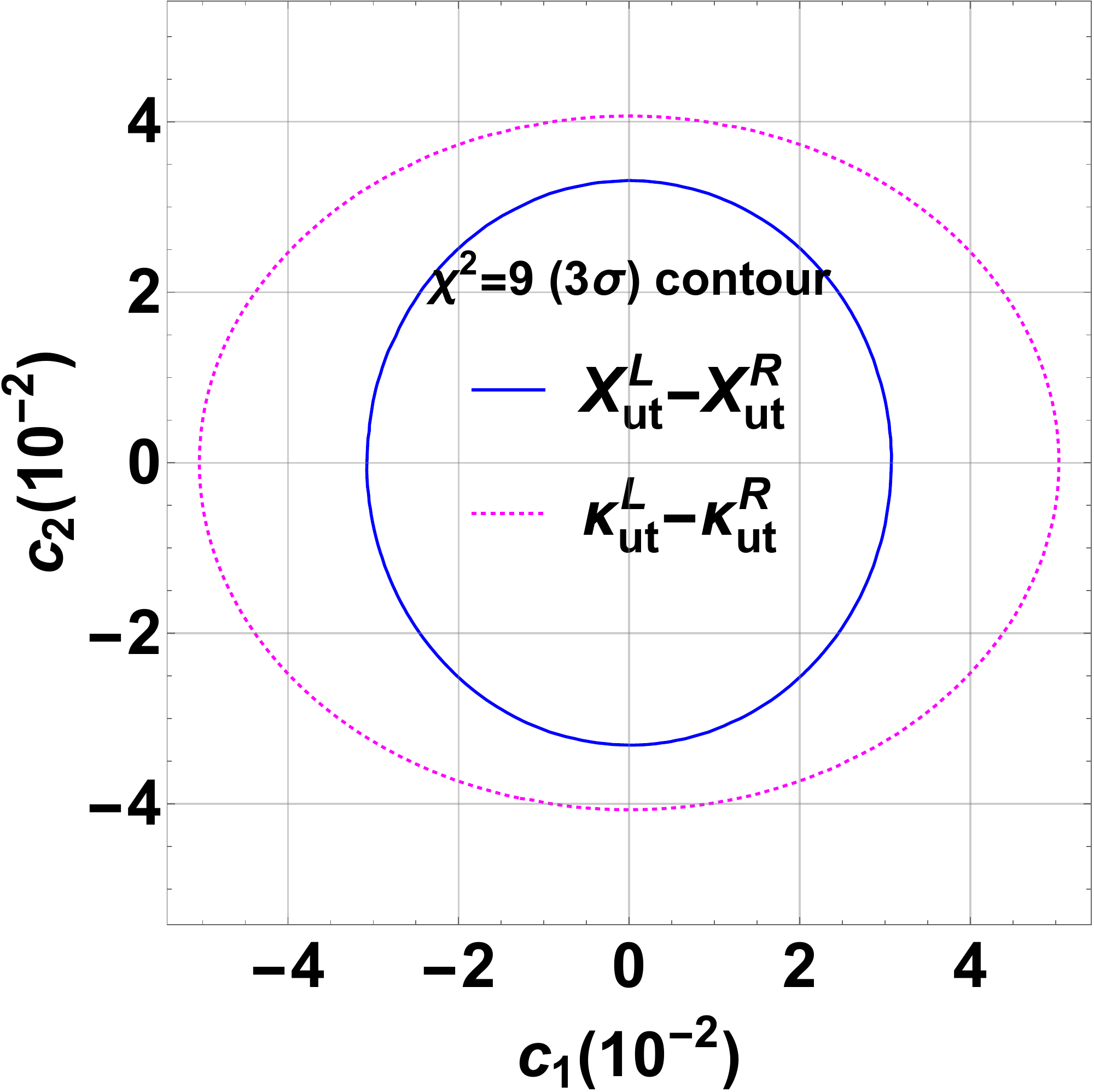}     
	\includegraphics[trim=0 0 0 0,clip,width=0.48\linewidth]{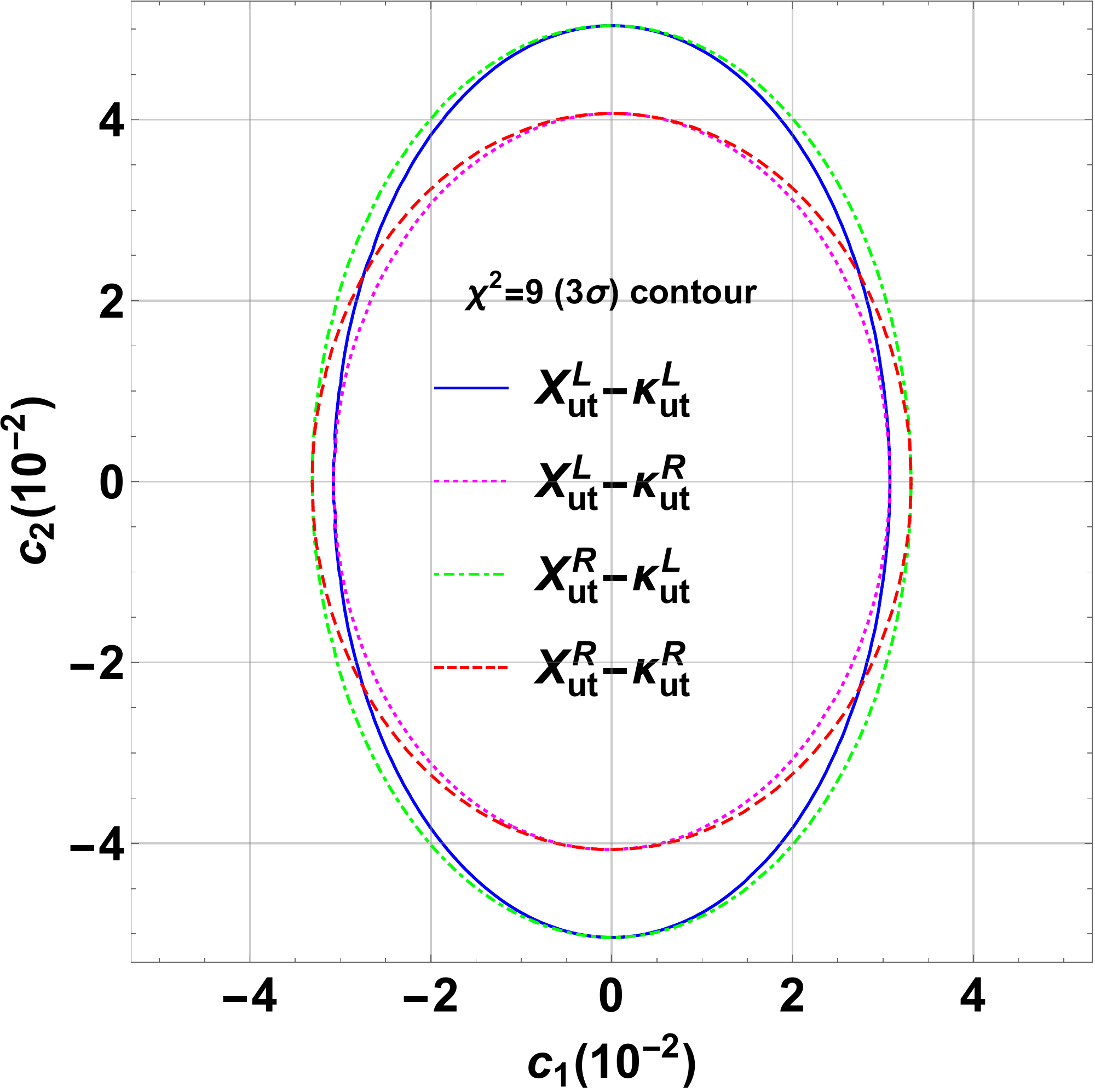} \\
	\includegraphics[trim=0 0 0 0,clip,width=0.48\linewidth]{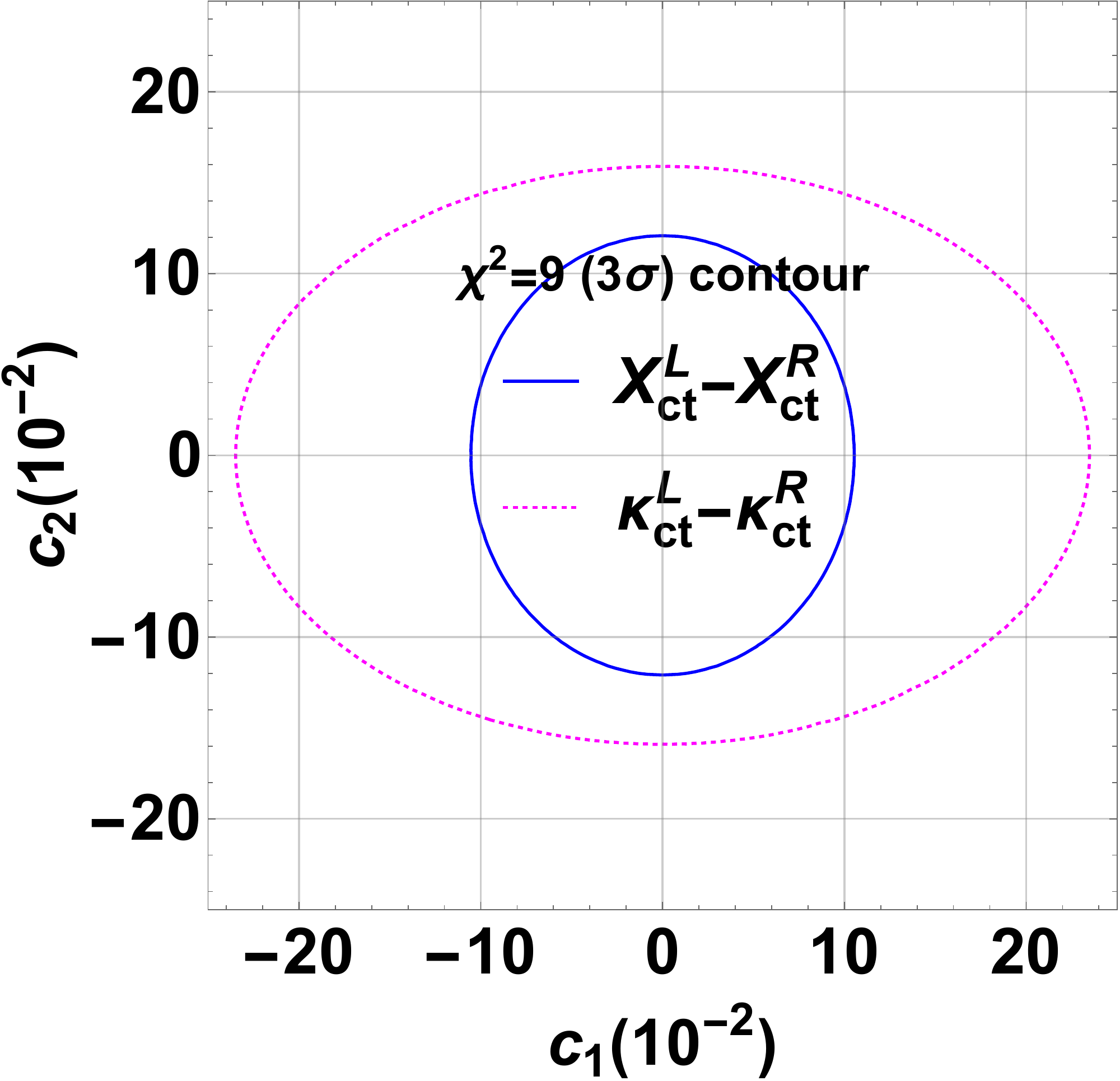}   
	\includegraphics[trim=0 0 0 0,clip,width=0.48\linewidth]{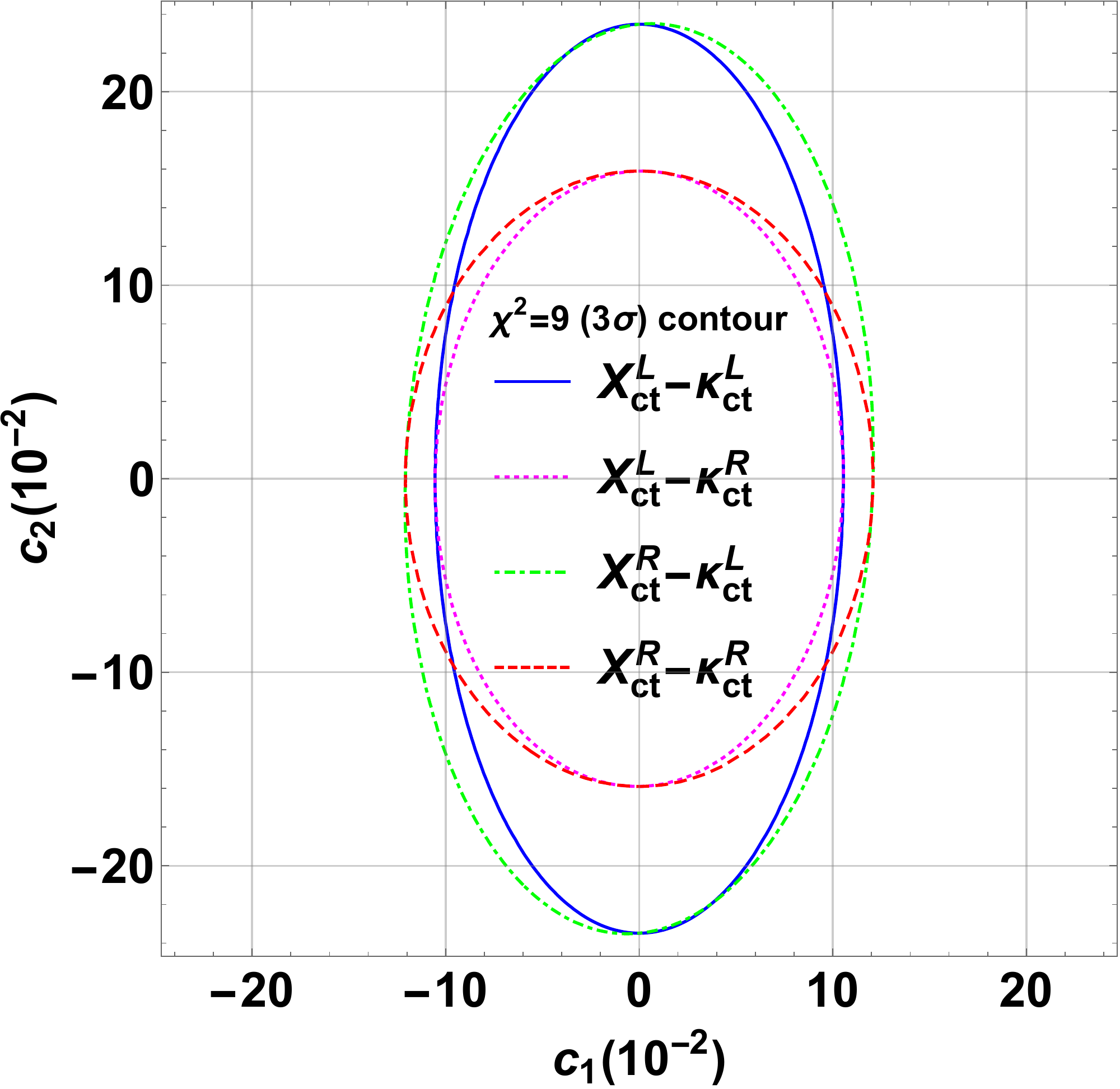}
	\caption{Two parameter reach with an integrated luminosity of 2~ab$^{-1}$. }
	\label{chi_two}
\end{figure}

We perform a $\chi^2$ analysis with the integrated cross section and the asymmetries considered as the observables ${\cal O}_i $, with
\begin{align}
  \chi^2\,({\bf f}) = \sum_i \frac{({\cal O}_i ({\bf f}))^2}
                      {\delta {\cal O}_i^2},
                      \label{eq_chi2}
\end{align}
where {\bf f} collectively denoting the anomalous couplings considered. $\delta {\cal O}$ is the estimated error in the measurement of ${\cal O}$, which is 
\(
\delta (\sigma) = \sqrt{\frac{\sigma_{BG}}{{\cal L}} + (\epsilon \sigma_{BG})^2}
\)
when cross section is considered as the observable, where $\sigma_{BG}$ denote the total background cross section (after final selection this is 2.3~fb in our case as shown in \cref{tab:cutflow_bkg}), and $\epsilon$ (taken to be $10\%$ in our numerical analysis) represents the systematic error in the calculation of cross section. When asymmetry is considered as the observable, we have \(\delta (A^i) = \sqrt{\frac{1-A^i_{BG}}{{\cal L}~\sigma_{BG}} + \epsilon^2_A}\), where $A^i_{BG}$ is the corresponding asymmetry arising purely from the background (once again, this is 2.3~fb in our case), and $\epsilon_A$ (again, taken to be $10\%$ in our numerical analysis) represents the systematic error in the calculation of asymmetries. Taking the cross sections $\sigma_{\rm tot}$ and asymmetries, $A_{x}, A_{z}$ and $A_{e}^{FB}$ as our observables we compute $\chi^2$ for single and multi-parameter cases. To distinguish the use of asymmetries used in this analysis in setting the limits of various couplings, we did the $\chi^2$ analysis with $\sigma$ as the only observable, which is represented in Fig.~\ref{chi_one} and Fig.~\ref{chi_two}. The single parameter case for an integrated luminosity of 2~ab$^{-1}$ is presented in Fig.~\ref{chi_one}. The vector couplings for the $Ztu$ case yield a $3\sigma$ limit of about $\pm 0.032~(\pm 0.034)$, whereas for the right-handed (left-handed) tensor couplings it is $\pm 0.042~(\pm 0.052)$. The corresponding values in the case of $Ztc$ couplings are $\pm 0.11~(\pm 0.12)$ and $\pm 0.16~(\pm 0.24)$ for the right-handed (left-handed) vector and tensor couplings, respectively. In \cref{chi_two}, we perform the same exercise for two parameter case.

\subsubsection{Likelihood mapping of the parameter space}
\label{sec:likelihood}
\begin{figure}[ht]
	\centering
	\includegraphics[trim=0 0 0 0,clip,width=0.48\linewidth]{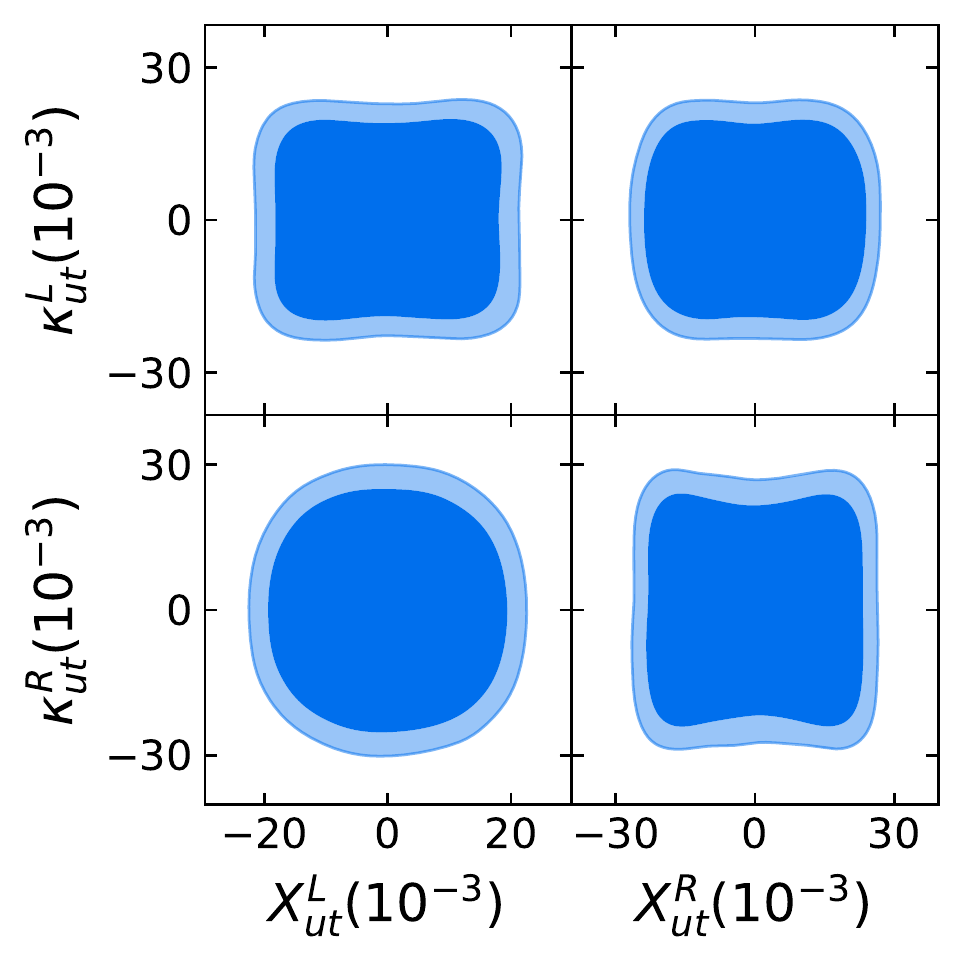} 
	\includegraphics[trim=0 0 0 0,clip,width=0.48\linewidth]{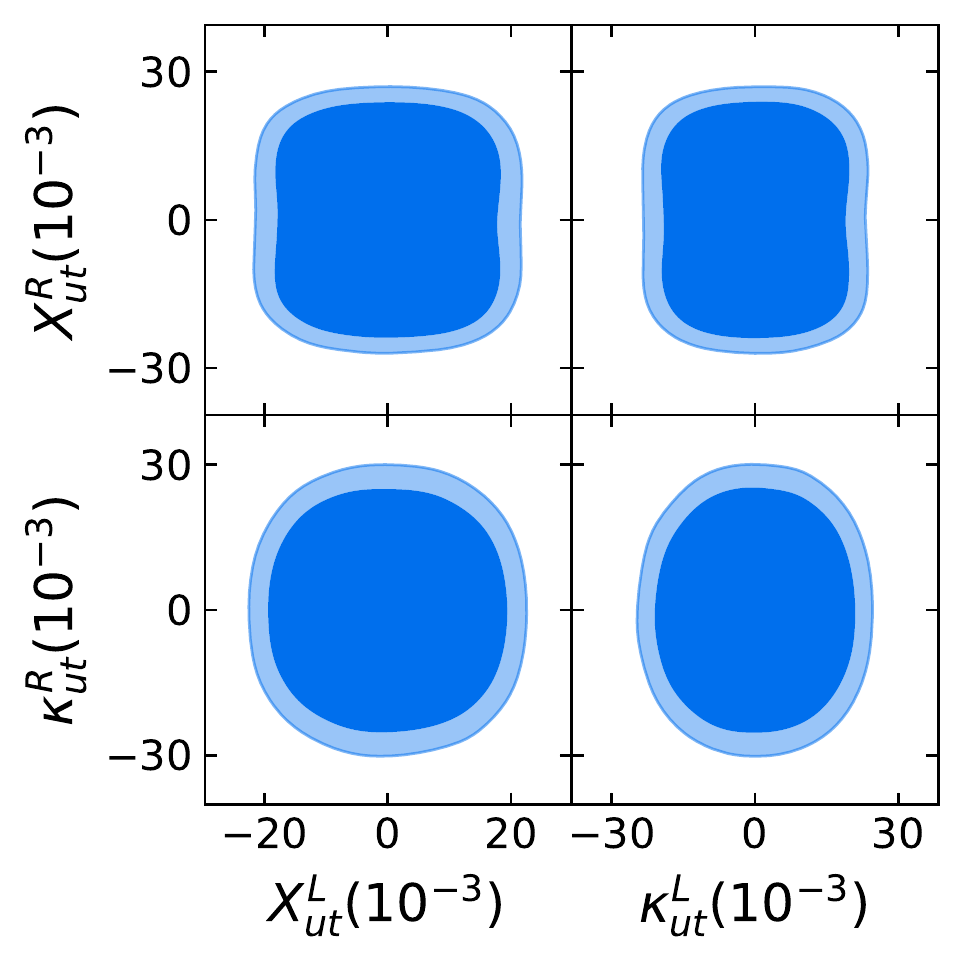} \\
	\includegraphics[trim=0 0 0 0,clip,width=0.48\linewidth]{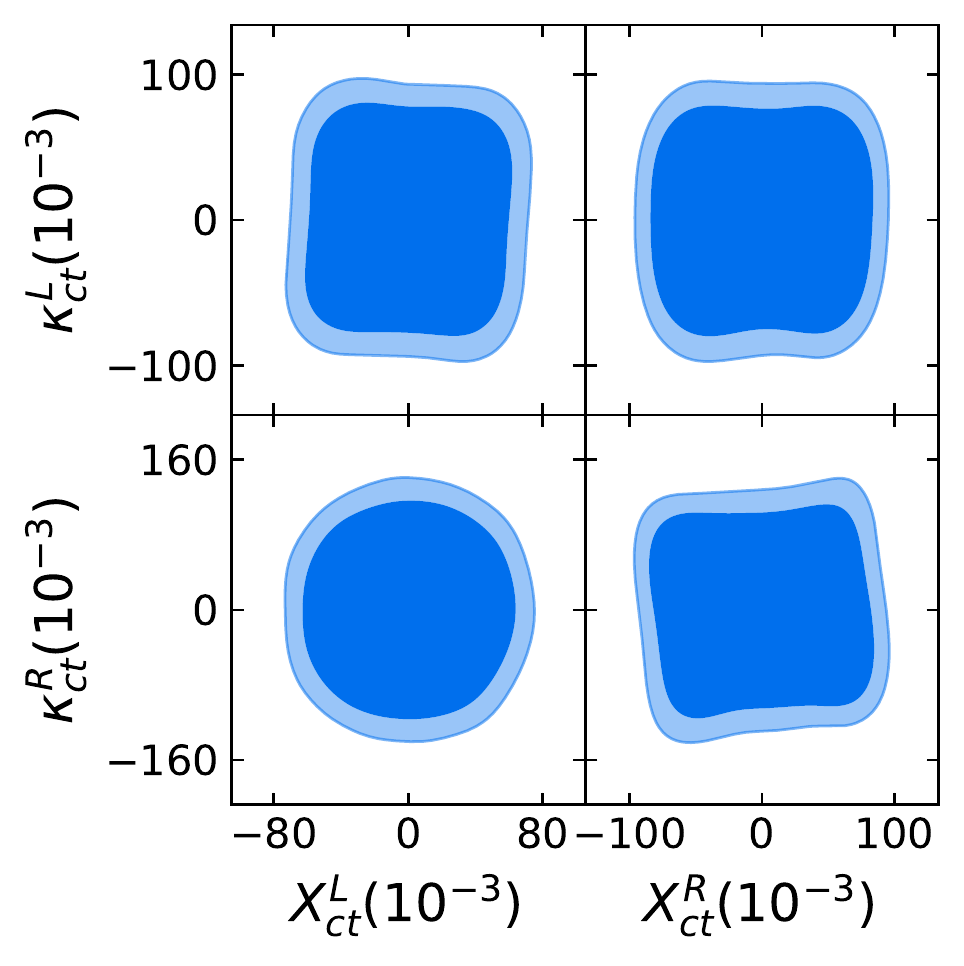} 
	\includegraphics[trim=0 0 0 0,clip,width=0.48\linewidth]{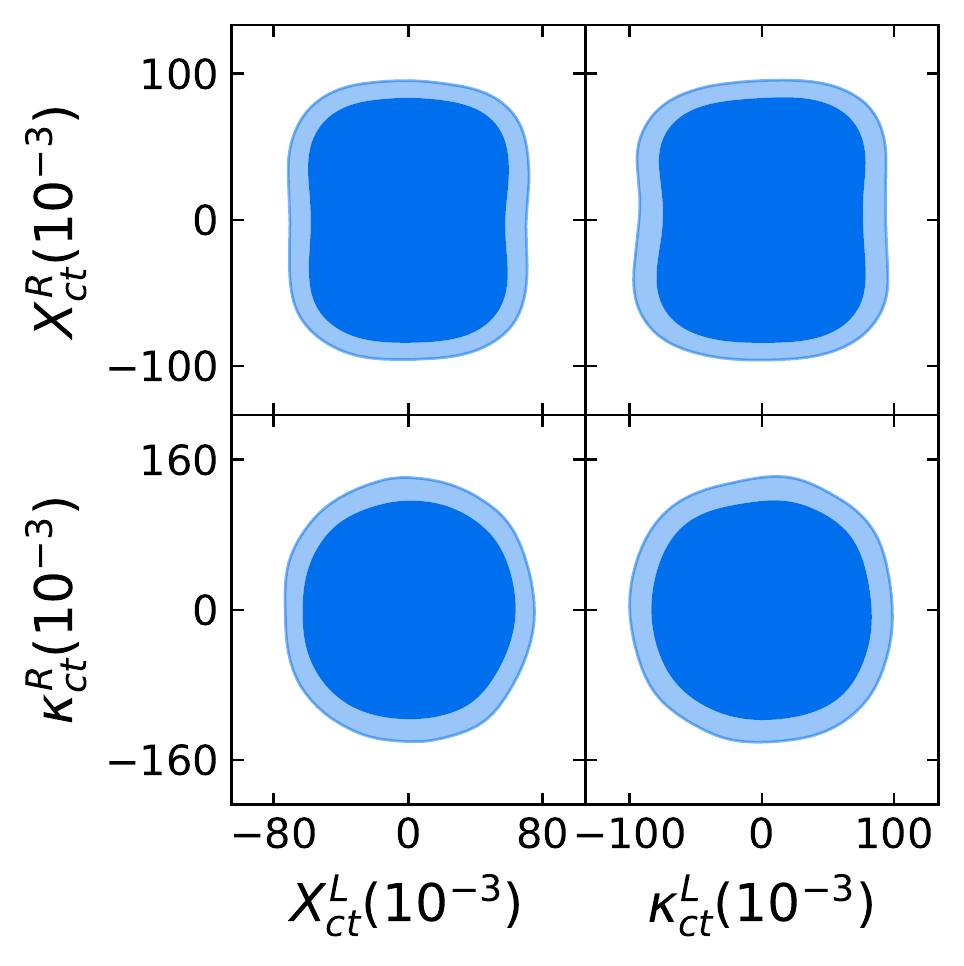}
	\caption{MCMC two parameter contour at 95\% (deep blue), 99\% (light blue) C.L.
		integrated luminosity of 2~ab$^{-1}$ is used. The tensor couplings are considered at $\Lambda=m_t$.}
	\label{fig:simul-contour}
\end{figure}

In this section we perform a likelihood analysis using the events available after the final selection. The likelihood of a given point ${\bf f}$ in the parameter space is given by
\begin{align}
 {\cal L}
  =
  \exp\left[ - \frac{\chi^2\,({\bf f})}
                             {2} \right]
\end{align}
where $\chi^2$ is defined in Eq.~\ref{eq_chi2}. We apply the Markov Chain Monte Carlo (MCMC) method to map the likelihood of the parameter space for each of the couplings. We make use of the publicly available {\sc GetDist}~\cite{Lewis:2013hha} package to obtain the single and multi-parameter bounds using MCMC chain. \cref{tab:simul-Limit_2000_delphes} shows the simultaneous limits on the anomalous couplings at 68\%, 95\% and 99\% C.L's obtained from the MCMC analysis considering an integrated luminosity of 2~ab$^{-1}$. 
\begin{table}[!ht]
	\centering
	{	\small 
		\begin{tabular} {c |c| c| c}
			\hline \hline
			Vector&\multicolumn{3}{c}{Obtainable reach (in part of $10^3$)} \\ \cline{2-4}
			Coupling         &at C.L. = 68\% & 95\% & 99\% \\\hline
			$X_{ut}^L               $ & $\in[   -9.8,  9.7 ]$   & $\in[    -16.4,  16.4  ]$  & $\in[    -20.0,  20.1]$    \\
			$X_{ut}^R              $ & $\in[  -13.7,  13.6   ]$   & $\in[   -20.8,  20.9   ]$  & $\in[   -24.4,  24.4]$    \\
			$X_{ct}^L               $ & $\in[   -30.8,  31.1  ]$   & $\in[  -52.8,  52.7]$  & $\in[  -65.7,  65.5]$    \\
			$X_{ct}^R              $ & $\in[ -48.0,  47.3]$   & $\in[  -73.0,  72.7]$  & $\in[  -85.9,  86.8]$    \\
 \hline \hline
		\end{tabular}}
\vskip 5mm
{\small		
		\begin{tabular} {c |c| c| c}
			\hline \hline
			Tensor&\multicolumn{3}{c}{Obtainable reach (TeV $^{-1}$)} \\ \cline{2-4}
			Coupling         &at C.L. = 68\% & 95\% & 99\% \\\hline
			$\kappa_{ut}^L  / \Lambda    $ & $\in[   -0.06, 0.06   ]$   & $\in[   -0.10,  0.10   ]$  & $\in[   -0.13,  0.13]$    \\
			$\kappa_{ut}^R /\Lambda     $ & $\in[   -0.07,  0.07  ]$   & $\in[   -0.12,  0.12   ]$  & $\in[   -0.16,  0.15]$ \\
			$\kappa_{ct}^L  /\Lambda   $ & $\in[ -0.23,  0.23$   & $\in[  -0.40,  0.40]$  & $\in[  -0.50,  0.50]$    \\
			$\kappa_{ct}^R /\Lambda     $ & $\in[ -0.31,  0.31  ]$   & $\in[  -0.57,  0.56]$  & $\in[  -0.71,  0.72]$    \\ \hline\hline
		\end{tabular}	}
		\caption{The list of simultaneous limits  on FCNC parameters obtained
	    from MCMC analysis including the cross-section and all other asymmetries for $e^-p$
		collider at $E_{e(p)} = 60~(7000)$~GeV with integrated luminosity of 2~ab$^{-1}$.}
	\label{tab:simul-Limit_2000_delphes}
\end{table}
For direct comparison with the experimental observations,  95\% branching fraction of FCNC decays of the top quark corresponding to the couplings quoted in Table~\ref{tab:simul-Limit_2000_delphes} are given in Table~\ref{brreach}. While these limits are at best comparable to that of the HL-LHC reach \cite{ATL-PHYS-PUB-2016-019}, one may notice that our limits do not assume the absence of other couplings, unlike those quoted in the case of LH-LHC study. 
\begin{table}[!ht]
	\centering
		\begin{tabular}{c|c|c|c} \hline
&BR \% &&BR \%  \\
&($t\to Zu$)&&($t\to Zc$)	\\\hline \hline
$X_{ut}^L=0.016$	&	{0.009}&$X_{ct}^L=0.053$&{0.095} \\\hline
$X_{ut}^R=0.021$&{0.015}&$X_{ct}^L=0.073$&{0.181} \\\hline
{\small $\frac{\kappa_{ut}^L}{\Lambda}=0.10$ TeV$^{-1}$}&{0.004}&{\small $\frac{\kappa_{ct}^L}{\Lambda}=0.40$ TeV$^{-1}$}&{0.068} \\\hline
{\small $\frac{\kappa_{ut}^R}{\Lambda}=0.12$ TeV$^{-1}$}&{0.006}&{\small $\frac{\kappa_{ct}^R}{\Lambda}=0.57$ TeV$^{-1}$}&{0.133} \\\hline
\end{tabular}
\caption{Limiting values of the couplings that can be reached (refer Table~\ref{tab:simul-Limit_2000_delphes}), and the branching fractions of the corresponding top quark decays.}
\label{brreach}
\end{table}
In \cref{fig:simul-contour} the 2-dimensional projections of the hyperspace of the 4-dimensional (four couplings for $u/c$ quark each) parameter space region limited by the 95\% and 99\% C.L. regions obtained assuming an integrated luminosity of 2~ab$^{-1}$. Limits of the order of 0.02 can be reached at 99\% C.L. on all the $Ztu$ couplings, whereas these are around 0.06-0.12 for the $Ztc$ couplings. 


\section{Discussion and conclusion}
\label{sec:conc}
Top quark, being the only quark that reveal direct information of the weak couplings through its decay without adding complication of hadronic bound states, has a special place to play in our understanding of the elementary particle dynamics. The fact that it is the heaviest of the known fundamental particles, consequently having the strongest coupling with the Higgs boson, makes the properties of the top quark a unique window to understand the electroweak symmetry breaking mechanism. In this article we have focused our attention on the FCNC couplings of the top quark with the $Z$ boson. Noting that the SM prediction of these couplings are about $10^{-15}$ to $10^{-12}$ orders of magnitude smaller than the present experimental bounds, there is ample opportunities in probing new physics effects in these couplings. However, extracting information regarding the FCNC couplings of top quark in the standard processes involving its rare decay (enabled by the new couplings) has limitations of statistics, as the decays are expected to be a few in a million of the standard decay process at the best. On the other hand, we may look at possibilities in the rare production, which are not possible in the absence of the FCNC couplings. Such rare single top productions at the LHC are harder to probe, as the standard single production processes overshadow these rare processes. Colliders with leptonic initial state like the electron-proton collider (LHeC) has definite advantage here, where it is possible to have $e^-p \rightarrow e^-t$ produced in the presence of $Ztq$ vertex, where $q$ denotes either $u$ or $c$ quark. There is no SM analogue of this process, and therefore expected to be relatively free from the background. We consider such a situation in a projected $e^-p$ collider of beam energies of 60 GeV (electron) and 7 TeV (proton) equivalent to a centre of mass energy of 1.3 TeV. We note that the presence of scattered (spectator) electron in this case is quite advantageous, and discuss exploiting it to distinguish the Lorentz structure of the anomalous coupling. We consider the leptonic decay of the top quark, and define angular asymmetries of the decay lepton that can be easily constructed. These angular asymmetries reflect the polarisation state of the top quark decayed. We have given a detailed discussion on the top-quark polarisation states, and the transition of the spin information to the decay leptons. Along with the integrated cross section, these additional observables are made use of in extracting the reach of the couplings. Anticipating small cross section as expected from couplings of the order of $10^{-3}$, we require large luminosity in the inverse attobarn range to have sufficient statistics for this investigation. Apart from single parameter analysis considering the presence of one coupling alone at a time, we perform a multi-parameter analysis, where $\chi^2$ minimisation and likelihood analysis methods are employed. MCMC technique is used for these analyses, with the cross section and asymmetries considered as observables. Correlations of the couplings in extracting the reach is obtained in a 4-dimensional hyperspace of the parameters, and the 2-dimensional slices of this in all combinations of two-parameter plane are presented. At an integrated luminosity of 2~ab$^{-1}$ we consistently obtain a reach of ${\cal O} (10^{-2})$ in the case of $Ztu$ and $Ztc$ vector couplings by both $\chi^2$ analysis and MCMC analysis at LHeC of $\sqrt{s}\approx 1.3$~TeV, whereas the reach of tensor couplings are about $0.1-0.5$ TeV$^{-1}$. While these limits sound approximately similar to that are projected in the case of HL-LHC (which assumes presence of single coupling at a time), note that our analysis considered the simultaneous presence of all the relevant couplings. We believe that this study has clearly brought out the advantages of the $e^-p$ collider in probing the top quark FCNC couplings with $Z$ boson, which would not only complement the information that could be extracted from the LHC, but also is capable of providing additional information like the Lorentz structure of the couplings.  Finally, we note that the scope of the present study is limited to the BSM scenarios leading to anomalous $Ztq$ vertices. Another class of models like the non-universal $Z'$ models \cite{He:2002ha,Langacker:2000ju, Fox:2011qd} with tree-level FCNC couplings could lead to the same process through $Z'$ exchange channels. Analysis of these models and a comparison with the framework considered in the present work is an interesting  project in itself.


\section*{Acknowledgements}
We acknowledge fruitful discussions within the LHeC Higgs group in customising the detector card used in this analysis. PP and RI thank the SERB-DST, India for the research grant EMR/2015/000333. PP, SB and RI thank the DST-FIST grant SR/FST/PSIl-020/2009 for offering the computing resources needed by this work. 

\bibliographystyle{apsrev4-1}
\bibliography{biblio}

\end{document}